\documentclass[]{spie}  
\usepackage[]{graphicx}
\usepackage[breaklinks,colorlinks,urlcolor=blue,linkcolor=blue,citecolor=blue]{hyperref}
\usepackage{color}

\title{A Transmissive X-ray Polarimeter Design For Hard X-ray Focusing Telescopes} 

\author{Hong Li\supit{a}, Hua Feng\supit{a}, Jianfeng Ji\supit{a}, Zhi Deng\supit{a}, Li He\supit{a}, Ming Zeng\supit{a}, Tenglin Li\supit{a}, Yinong Liu\supit{a},
Peiyin Heng\supit{a}, Qiong Wu\supit{a}, Dong Han\supit{a}, Yongwei Dong\supit{b}, Fangjun Lu\supit{b}, Shuangnan Zhang\supit{b}
\skiplinehalf
\supit{a}Department of Engineering Physics and Center for Astrophysics, Tsinghua University, Beijing 100084, China; \\
\supit{b}Key Laboratory of Particle Astrophysics, Institute of High Energy Physics, Chinese Academy of Sciences, Beijing 100049, China; \\
}

\begin{document} 
\maketitle 

\begin{abstract}
The X-ray Timing and Polarization (XTP) is a mission concept for a future space borne X-ray observatory and is currently selected for early phase study. We present a new design of X-ray polarimeter based on the time projection gas chamber. The polarimeter, placed above the focal plane, has an additional rear window that allows hard X-rays to penetrate (a transmission of nearly 80\% at 6 keV) through it and reach the detector on the focal plane. Such a design is to compensate the low detection efficiency of gas detectors, at a low cost of sensitivity, and can maximize the science return of multilayer hard X-ray telescopes without the risk of moving focal plane instruments. The sensitivity in terms of minimum detectable polarization, based on current instrument configuration, is expected to be 3\% for a 1mCrab source given an observing time of $10^5$~s. We present preliminary test results, including photoelectron tracks and modulation curves, using a test chamber and polarized X-ray sources in the lab. 
\end{abstract}


\section{INTRODUCTION}
\label{sec:intro}  

In the X-ray band, typically at a few keV, high degree of linear polarization is expected from high energy astrophysical objects where magnetic field or asymmetric geometry plays a key role in radiation or radiation transfer \cite{Kallman2004}. X-ray polarimetry can help measure the magnetic field in supernova remnants and constrain the particle acceleration models \cite{Reynolds1990}. One can test the emission mechanism for rotation-powered pulsars \cite{Daugherty1996, Muslimov2004, Romani1996}, distinguish the fan model and pencil model for strongly magnetized accreting pulsars \cite{Meszaros1988}, and measure the geometry for millisecond pulsars \cite{Viironen2004}, with phase resolved X-ray polarimetry. X-ray polarimetry for high synchrotron peaked blazars will be a powerful probe for the magnetic structure in relativistic jets \cite{Marscher2008}. It is also useful to constrain the accretion disk inclination, which is needed for an unbiased estimate for the black hole spin \cite{Li2009,Schnittman2009}, or test the symmetry for the Comptonization corona around X-ray binaries or active galactic nuclei \cite{Schnittman2010}. Sensitive X-ray polarimetry will open a new window in X-ray astronomy in addition to imaging, spectroscopy and timing \cite{Soffitta2013}.

Due to the short wavelengths, X-rays are free of Faraday rotation during propagation, which allows us to probe the innermost region where the emission arises. In the past, measurement of X-ray polarization is usually based on Bragg diffraction or Thomson scattering, which is either limited to a narrow band and/or of poor efficiency. To date, the only successful experiment was made in 1970s using mosaic graphite crystals onboard the OSO-8 satellite. It obtained the first firm and positive result for the Crab nebula \cite{Weisskopf1976, Weisskopf1978}, zero or low time-averaged polarization for Sco X-1 and Cyg X-1 \cite{Long1979,Long1980}, and loose upper limits for other sources \cite{Hughes1984}. High sensitivity X-ray polarimetry still remains unexplored in astronomy due to the lack of technology.  

Recently, with the development of micro-pattern gas detectors (MPGDs), sensitive X-ray polarimetry becomes possible by means of measuring photoelectron tracks following the absorption of X-rays in a gas detector \cite{Costa2001, Bellazzini2003}. This is based on the fact that the emission angle of the photoelectron is dependent on the polarization angle of the incident photon. On the plane perpendicular to the photon direction, the emission angle for photoelectrons follows a $\cos^2$ distribution.  The electron track length is on the order of millimeter, and only the initial part before significant scattering happens is useful for polarization measurement.  MPGDs like the gas electron multiplier (GEM) detector with a position resolution close to 100 $\mu$m was proven to be able to resolve the electron track and measure the X-ray polarization. Mission concepts based on such techniques including the Gravity and Extreme Magnetism SMEX (GEMS) \cite{Swank2010} and the X-ray Imaging Polarimetry Explorer (XIPE) \cite{Soffitta2013} were proposed. The techniques for GEMS and XIPE are different. XIPE uses pixel detectors that can directly image the photoelectron, which is intrinsically of low systematics \cite{Costa2001, Bellazzini2003, Bellazzini2004, Bellazzini2006a, Bellazzini2006b, Muleri2008, Bellazzini2010, Muleri2012}. GEMS uses one dimensional time projection chambers (TPCs) and the photoelectron image is measured by different means in X and Y (i.e., strip position and timing).  The TPC polarimeter may have a higher efficiency but requires a huge amount of effort for calibration in order to reach a low systematic error as for the imaging polarimeter. Thus, the two techniques may be used for different purposes. The imaging detector is suitable for extended sources like supernova remnants and bright objects with low polarization like the accreting X-ray binaries, while the TPC polarimeter is good for highly polarized but relatively faint objects.

The X-ray Timing and Polarization (XTP) was proposed as a future observatory-class space mission concept in China, with an unprecedented sensitivity in X-ray polarimetry. A lot of its key science objectives will rely on X-ray polarimetry or joint analysis with spectroscopy, timing and polarimetry.  The choice of polarization techniques is still under investigation. Both the imaging and TPC polarimeters are currently under development at Tsinghua University. In this paper, we will introduce a modified design of the TPC X-ray polarimeter based on the version for GEMS, which can compensate the low efficiency of gas detectors and maximize the utilization of the focal plane behind hard X-ray focusing telescopes. We will show the preliminary results obtained with a test chamber and the expected sensitivity in X-ray polarimetry. 

\section{Polarimeter}
\label{sec:design}

The optics of XTP includes multilayer coated mirrors, which are sensitive in the energy range up to 40 keV \cite{Jiang2014}. There is no movable part on the focal plane of XTP. Therefore, if a gas detector like a photoelectric X-ray polarimeter is mounted on the focal plane of high energy optics, it will be a huge waste for the high energy photons. The TPC polarimeter with a long absorption depth may have a higher quantum efficiency than the imaging polarimeter. However, due to the choice of a low pressure and low-Z gas mixture, which is optimized for imaging the tracks for electrons of a few keV like the GEMS TPC polarimeter \cite{Black2010}, the stopping power for photons above 10 keV is still low.  

\begin{figure}[!htbp]
\begin{center}
\includegraphics[width=0.7\textwidth]{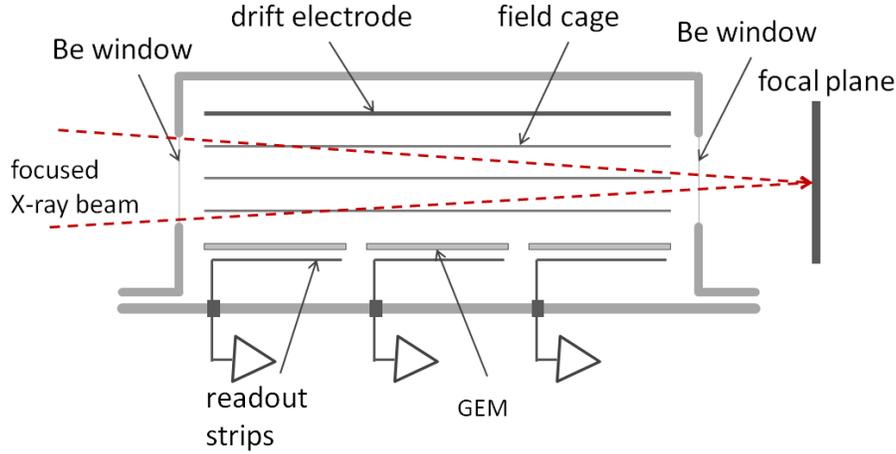}
\end{center}
\caption[det]
{\label{fig:det} 
Schematic drawing of the TPC polarimeter, not to scale.}
\end{figure}

To overcome this problem, we propose to open a second, rear window on the TPC polarimeter, and mount it above the focal plane. Thus, high energy photons that are mostly  unabsorbed in the chamber can penetrate through it and be collected by other focal plane detectors, e.g, semiconductor arrays. A schematic drawing of the polarimeter design is shown in Figure~\ref{fig:det}. As the focal length is about 3.5 m for XTP, the beam is quasi-parallel and the beam size is still small and close to the focal spot size if the chamber is placed above the focal plane. Along the light path, the sensitive volume is divided into several independent units, which are read out individually in order to increase the reliability and decrease the readout capacitance. Each unit consists of a GEM foil for signal amplification, one-dimensional strip electrodes for signal collection, and electronics including multi-channel high-frequency preamplifiers, shaping amplifiers, and a capacitor array for signal processing and storage. The cage enclosing the total sensitive volume is to shape a uniform electric field for electron drifting. 

For the XTP design, the polarimeter consists of three readout units. The GEM foils are 100$\mu$m thick insulated by liquid crystal polymer \cite{Tamagawa2009}. The holes are laser-etched with a diameter of 70~$\mu$m and a pitch of 140~$\mu$m in a hexagonal pattern. Three titanium plates serve as the field cage to shape a uniform electric field which is necessary for electron drifting.  The readout plate is a flexible printed circuit board containing 128 parallel strips with a pitch of 140~$\mu$m, matching the spacing of hole rows in the same direction. The total readout region is 1.8 cm wide and 8 cm long. The signals collected by the strips are fed into the electronics in the air side of the chamber via a feedthrough that seals the flexible board.  

\begin{figure}[!htbp]
\begin{center}
\includegraphics[width=0.6\textwidth]{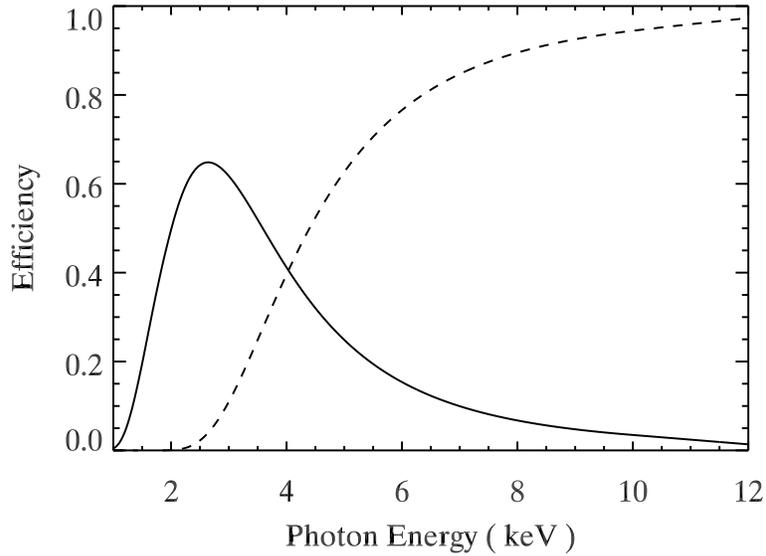}
\end{center}
\caption[eff]
{\label{fig:eff} 
The detection efficiency (solid line) and the total transmission (dashed line) for the transmissive polarimeter.}
\end{figure}

We adopt the pure dimethyl ether (DME) with a pressure of a quarter of atmosphere as the gas mixture. DME has the smallest transversal diffusion coefficient among commonly used gases for gaseous detectors \cite{Sharma1998}. Such a gas mixture has been proven viable for a TPC polarimeter \cite{Black2010,Hill2012}. With three readout units, as shown in Figure~\ref{fig:det}, the detector has a total length of 28 cm and a sensitive absorption depth of 24 cm. A beryllium foil of 50 $\mu$m is used as the entrance window and the back window is 100 $\mu$m thick because most soft photons are stopped by the gas.  With such a configuration, the detection efficiency and the transmission curves versus energy are shown in Figure~\ref{fig:eff}. The detection efficiency peaks at 2--3~keV, which is the peak of most high energy astrophysical objects after interstellar absorption. The transmission at 6.4 keV, the energy for neutral iron K$\alpha$ lines, is as high as 80\%.

\begin{figure}[!htbp]
\begin{center}
\includegraphics[width=0.45\textwidth]{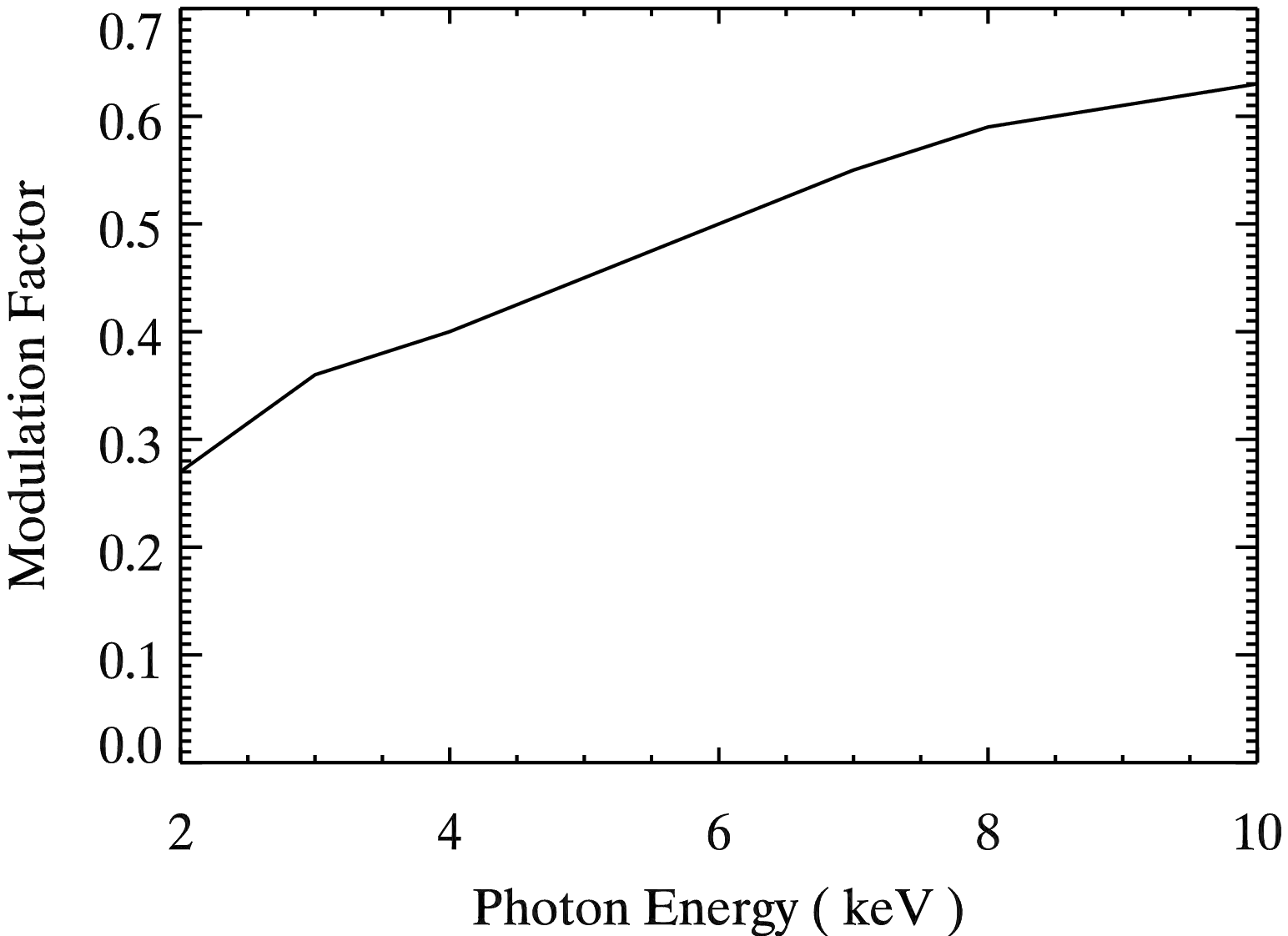}
\includegraphics[width=0.45\textwidth]{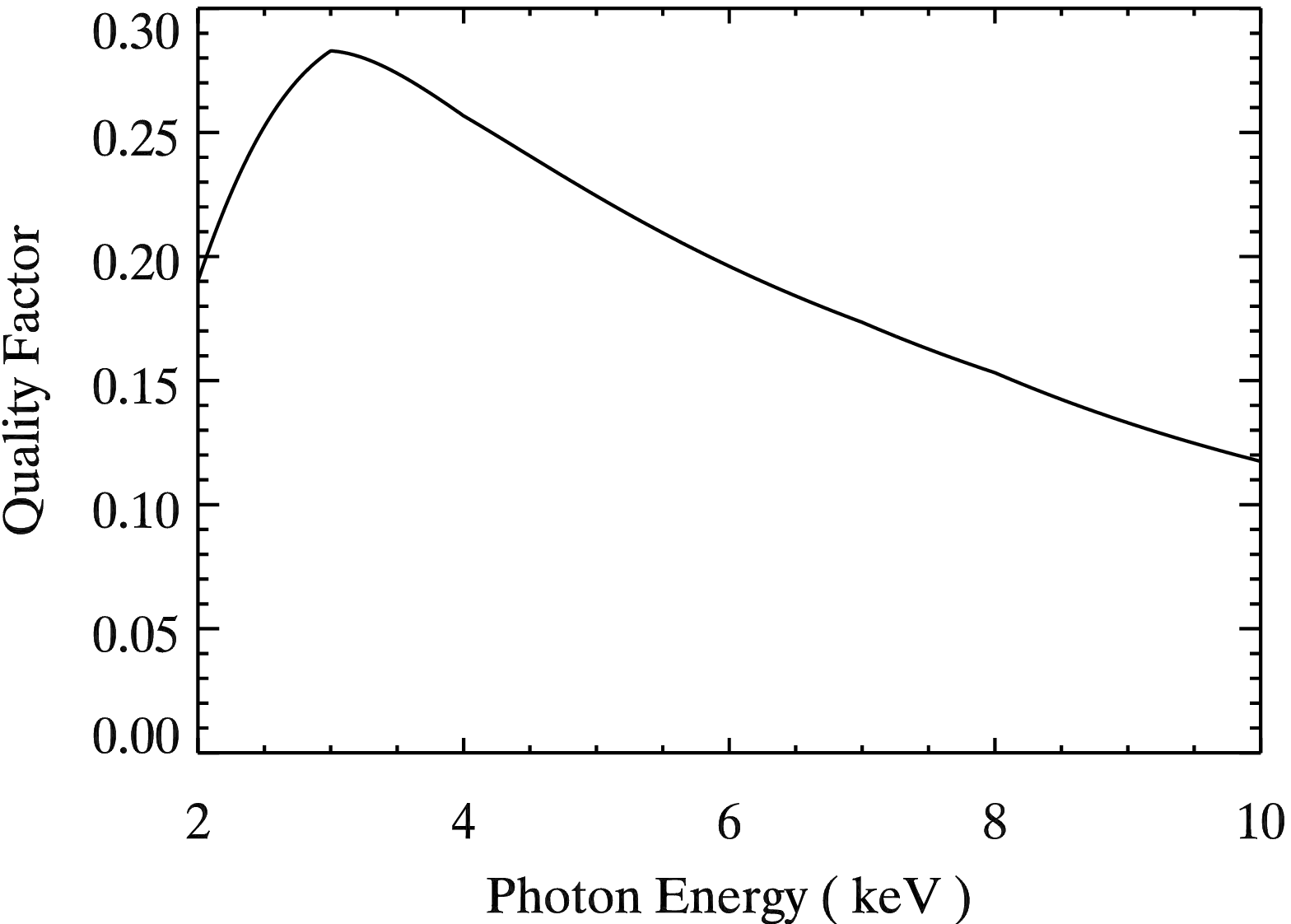}
\end{center}
\caption[qf]
{\label{fig:qf} 
Modulation factor (left) and quality factor (right) versus photon energy.}
\end{figure}

Using Geant4 simulations, and incorporating the transversal diffusion coefficient found by Magboltz and electronics noise from measurement, we simulated electron tracks and estimated their emission angles following the so-called ``impact point'' reconstruction algorithm \cite{Depaola2006}. The modulation factor $\mu$, defined as the fraction of modulation in response to a fully polarized source, is shown in Figure~\ref{fig:qf}. It gradually increases from nearly 0.3 to 0.6 with photon energies from 2 to 10 keV. The sensitivity of a polarimeter is scaled with the modulation factor times the square root of the detection efficiency, which is defined as the quality factor,  is also shown in the same figure. As one can see, the polarimeter is mostly sensitive in the energy band of 2--6 keV. 

\section{Test with a prototype}
\label{sec:result}

\begin{figure}[!htbp]
\begin{center}
\includegraphics[height=9cm]{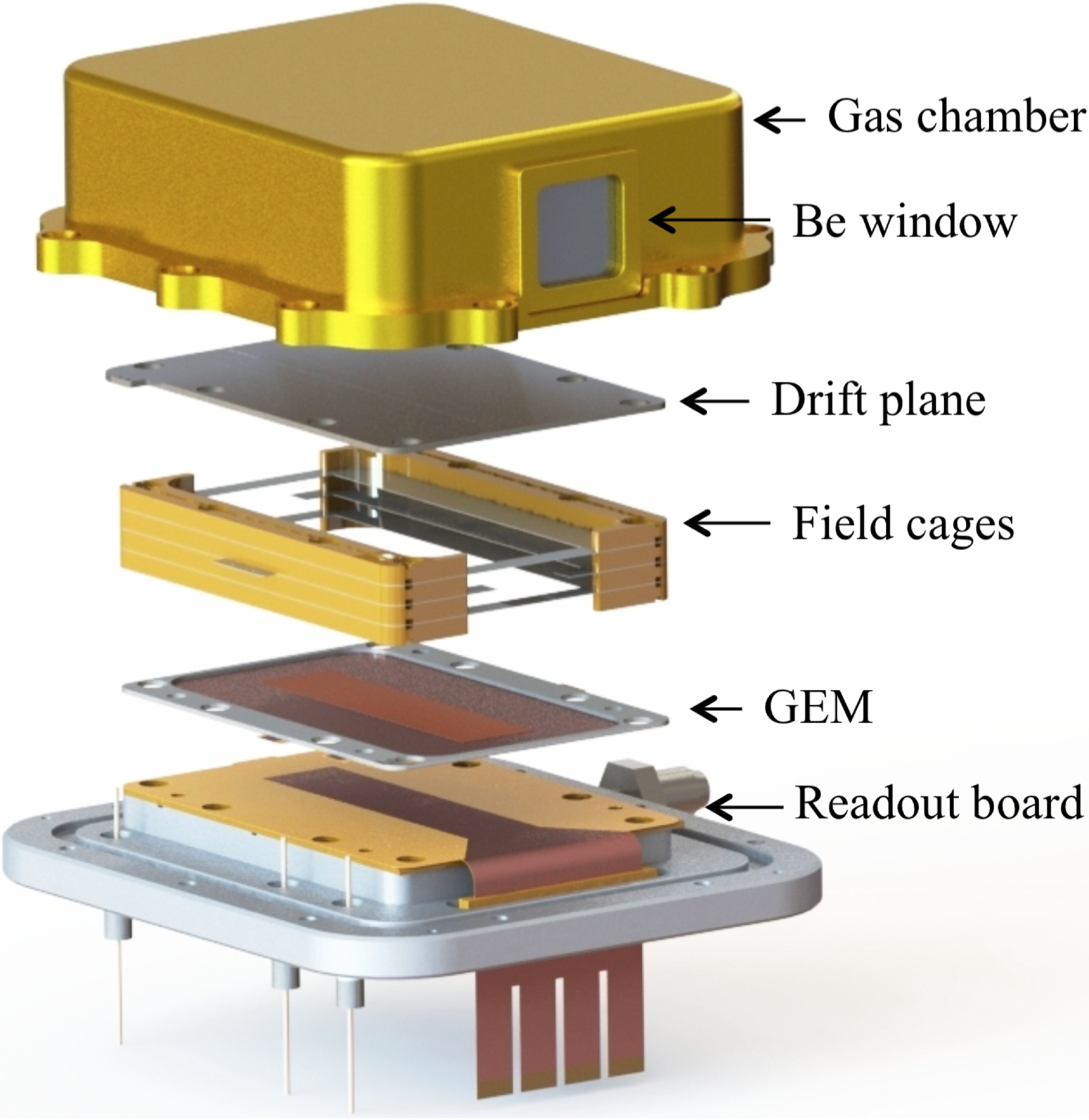}
\includegraphics[width=0.45\textwidth]{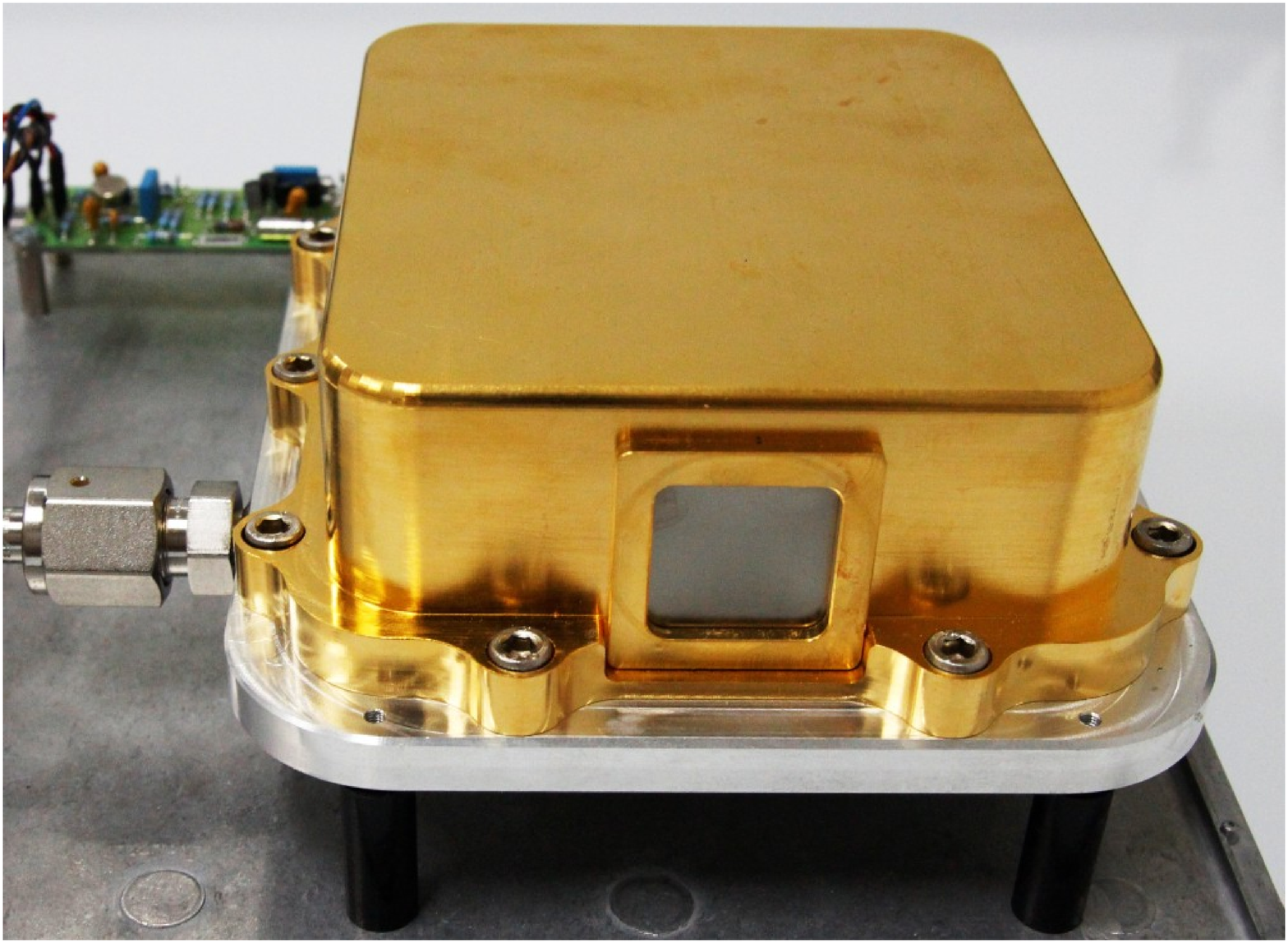}
\end{center}
\caption[prototype]
{\label{fig:prototype} 
Schematic drawing (left) and a picture (right) of the test prototype.}
\end{figure}

\begin{figure}[!htbp]
\begin{center}
\includegraphics[width=0.95\textwidth]{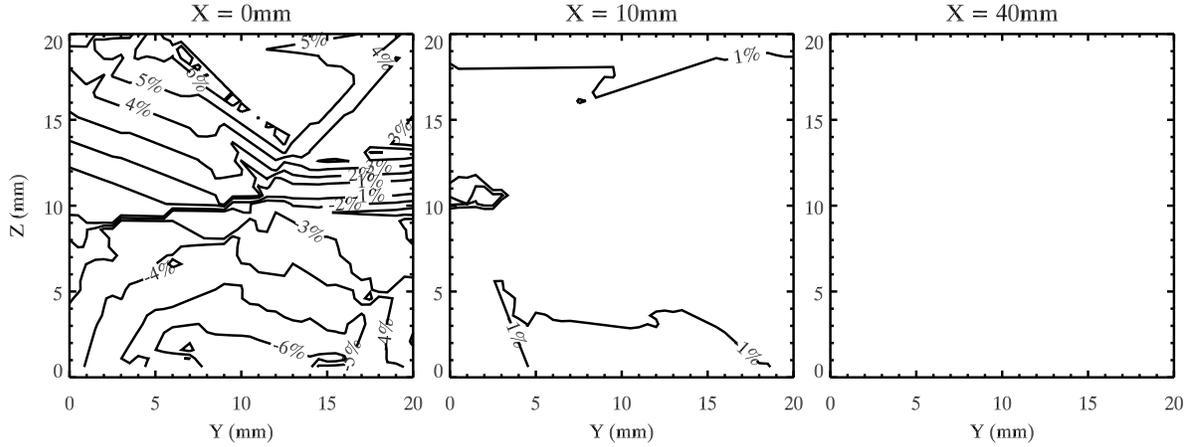}
\end{center}
\caption[efield]{
\label{fig:efield} 
Electric field contours at different cross-sections of the chamber. $X$ is the direction along the light path, and $Z$ is the electron drift direction.  The total sensitive volume has a length of 80~mm, i.e., from $X = 0$ to $X = 80$~mm. The nonuniformity of the filed is 4.1\% near the edge and 0.1\% at the center. }
\end{figure}

A test prototype for the polarimeter (see Figure~\ref{fig:prototype}) was manufactured and assembled by Oxford Instrument Analytical Oy.  Figure~\ref{fig:efield} shows the electric field uniformity at different cross-sections along the long direction of the sensitive region obtained by Maxwell. The field nonuniformity, defined as the standard deviation of the field strength divided by the mean value, is 4.1\% at the very front and back end of the chamber, and is 0.1\% at the center of the chamber along the light path. The readout electronics is being developed at Tsinghua University, which is a TPC chip with a peaking time of 50 ns and a sampling rate of 40 MHz. The current chip has only 16 channels, based on which we built a readout board with 32 channels with two of them. This only allows us to read out part of the collection region for functional test. A careful calibration can be done only if a full scale readout  system is available. 

\begin{figure}[!htbp]
\begin{center}
\includegraphics[width=0.25\textwidth]{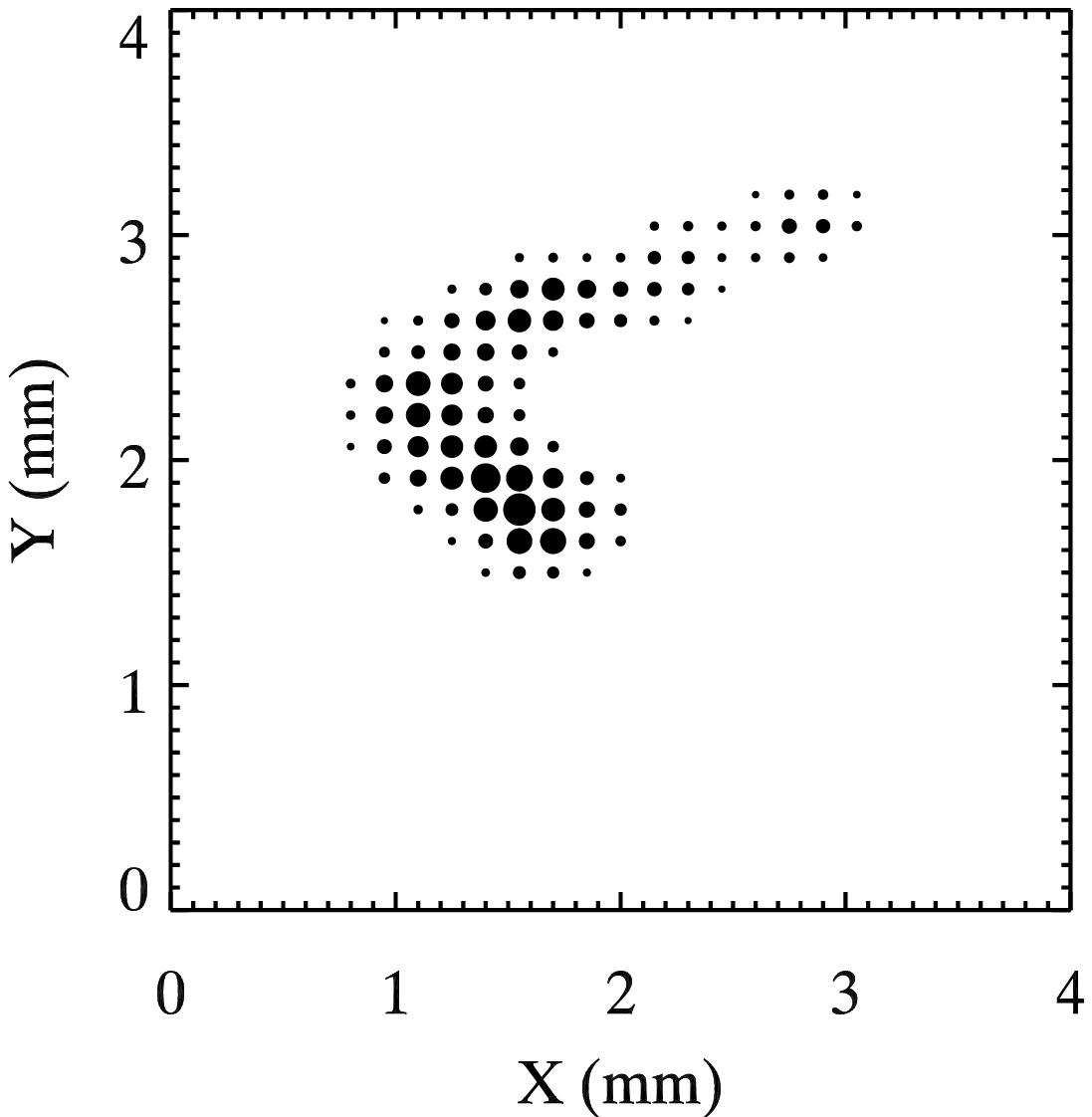}
\includegraphics[width=0.25\textwidth]{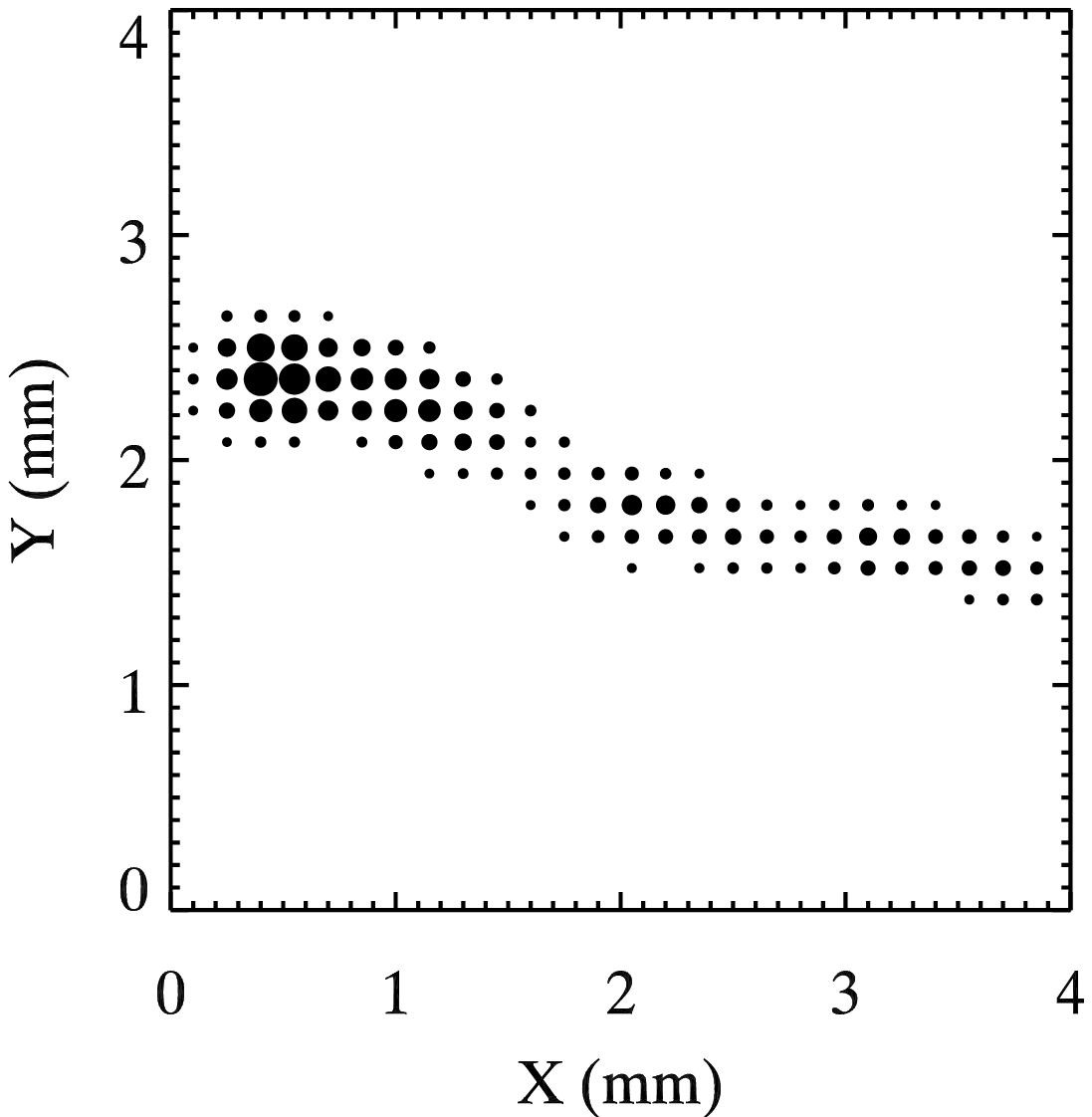}\\
\includegraphics[width=0.25\textwidth]{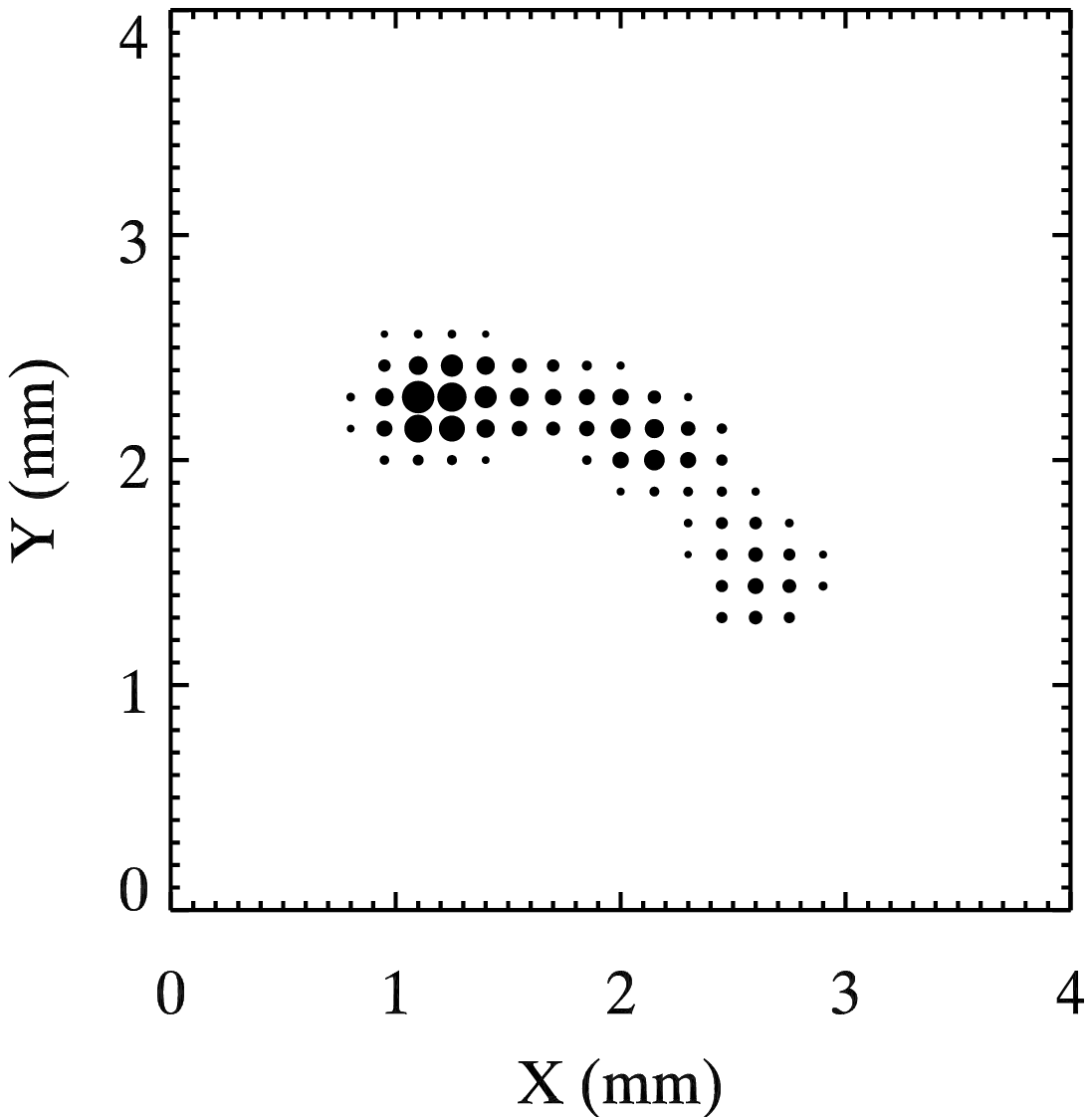}
\includegraphics[width=0.25\textwidth]{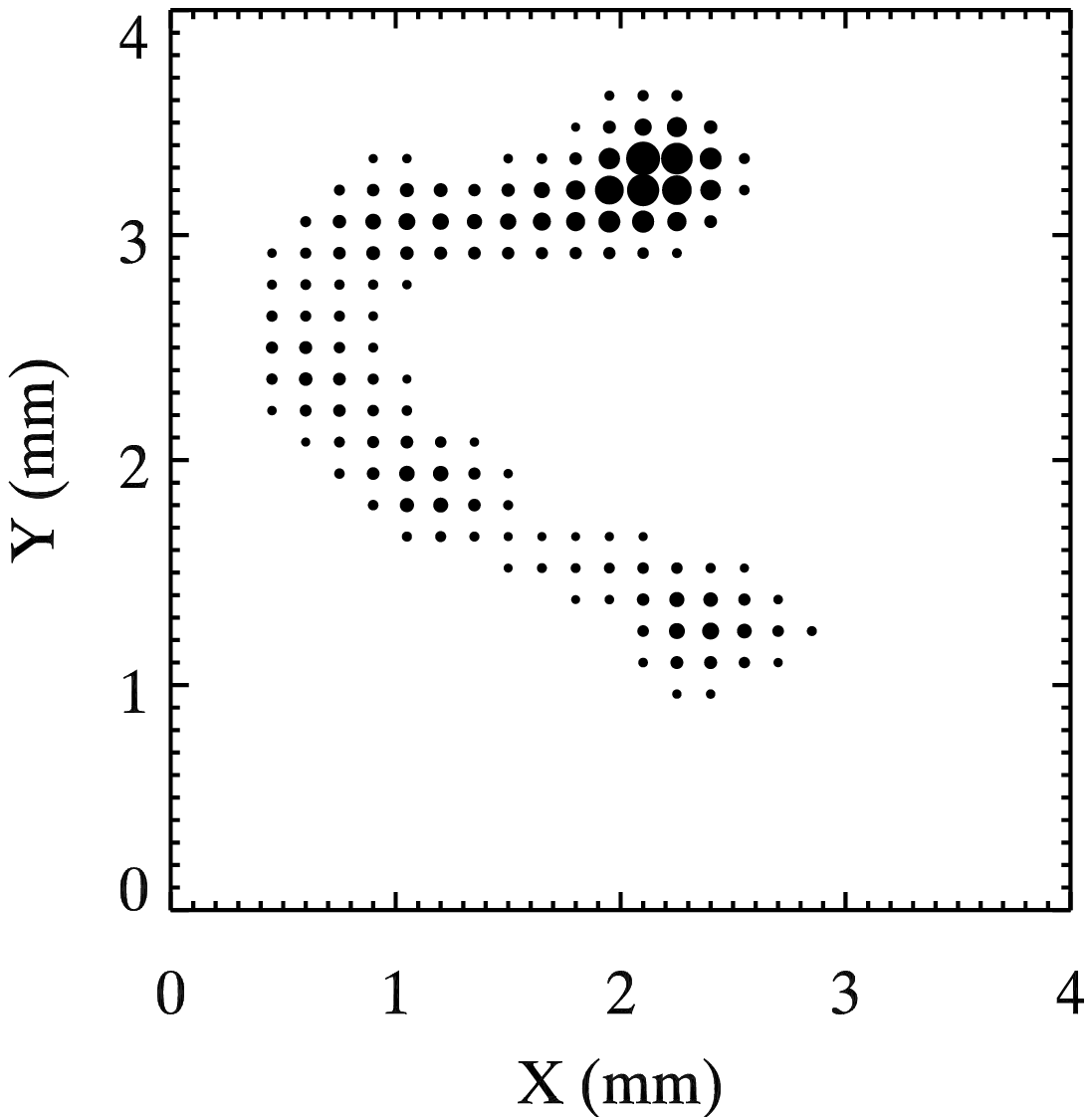}
\end{center}
\caption[track]
{\label{fig:track} 
Measured photoelectron tracks for 6.1~keV (left) and 8.0~keV (right) X-rays.}
\end{figure}

\begin{figure}[!htbp]
\begin{center}
\includegraphics[width=0.45\textwidth]{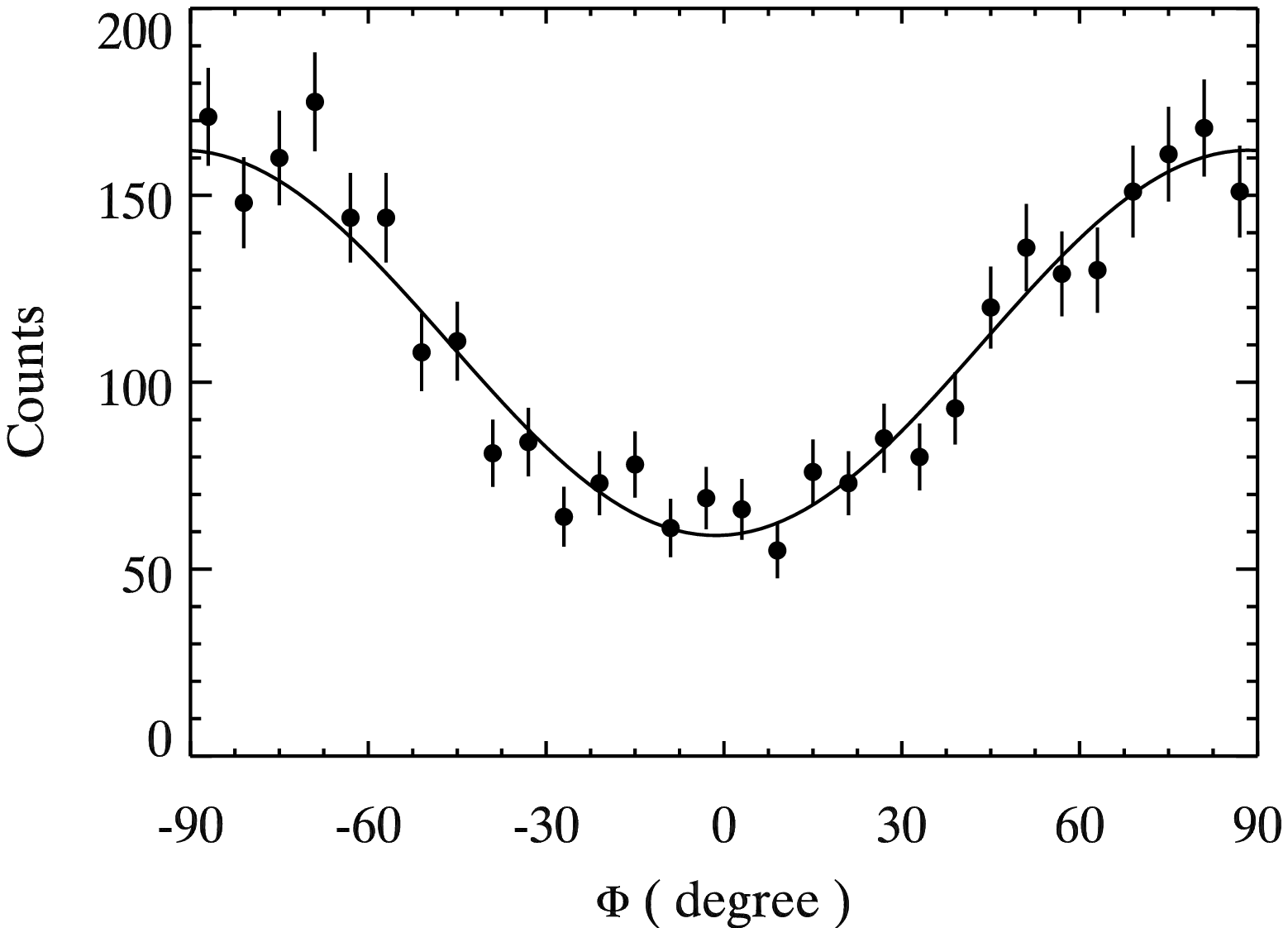}
\includegraphics[width=0.45\textwidth]{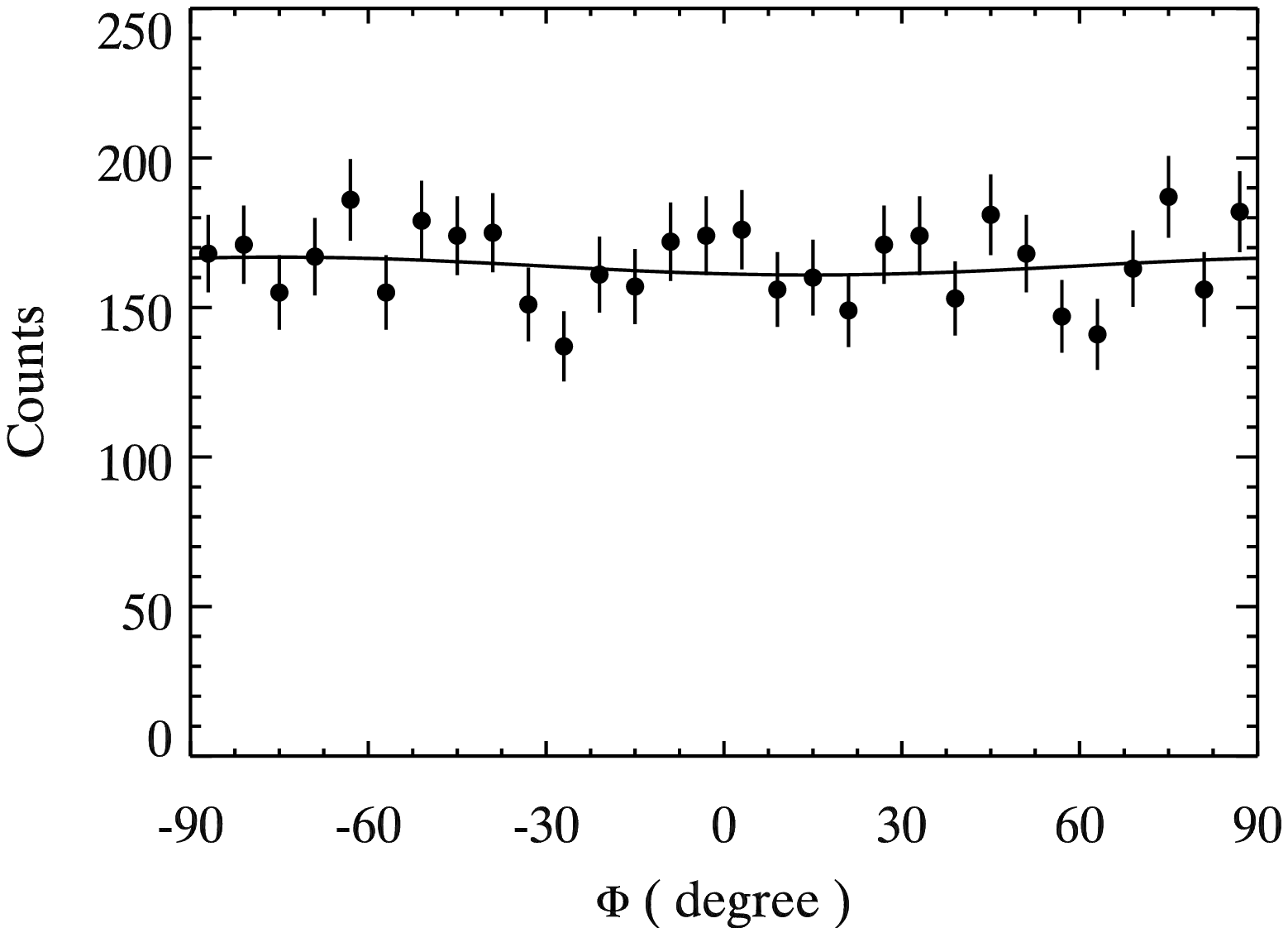}
\end{center}
\caption[mu]
{\label{fig:mu} 
Modulation curves obtained with a fully polarized beam at 6.1 keV (left) and an unpolarized beam at 8.0 keV (right). The modulation factor is measured to be $47\% \pm 2\%$ at 6.1 keV, while for the unpolarized beam, it is not quoted because the current setup has an incomplete readout. }
\end{figure}

A fully polarized source is produced by diffracting X-rays from a tube on a LiF crystal tilted at an angle of 45 degrees. The beam has an energy of 6.1 keV and is tested to be fully polarized using a second crystal at 45 degrees. The copper K$\alpha$ line at 8.0 keV illuminated by an X-ray tube was used as a non-polarized source. Figure~\ref{fig:track} shows some images of the photoelectron tracks measured with this detector. The modulation curves obtained with the fully polarized beam at 6.1 keV and the unpolarized beam at 8.0 keV are displayed in Figure~\ref{fig:mu}.  The modulation factor at 6.1 keV is measured to be $47\% \pm 2\%$ and is consistent with that from simulations.  We do not quote the modulation factor for the unpolarized beam because the current setup has only 32 channels and the limited width will cause some systematics more or less. 

\begin{figure}[!htbp]
\begin{center}
\includegraphics[width=0.6\textwidth]{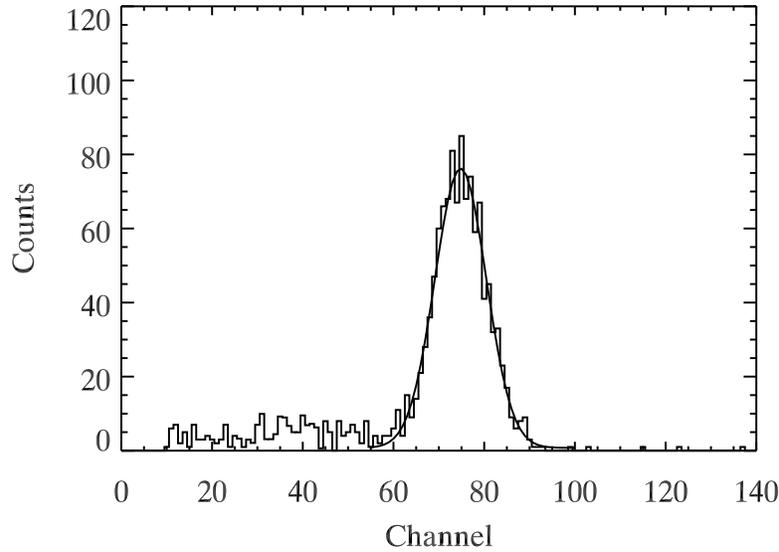}
\end{center}
\caption[spec]
{\label{fig:spec} 
Energy spectrum for X-rays at 6.1~keV. A Gaussian profile was adopted to fit the spectrum and the measured energy resolution is 17\%.}
\end{figure}

\begin{figure}[!htbp]
\begin{center}
\includegraphics[width=0.45\textwidth]{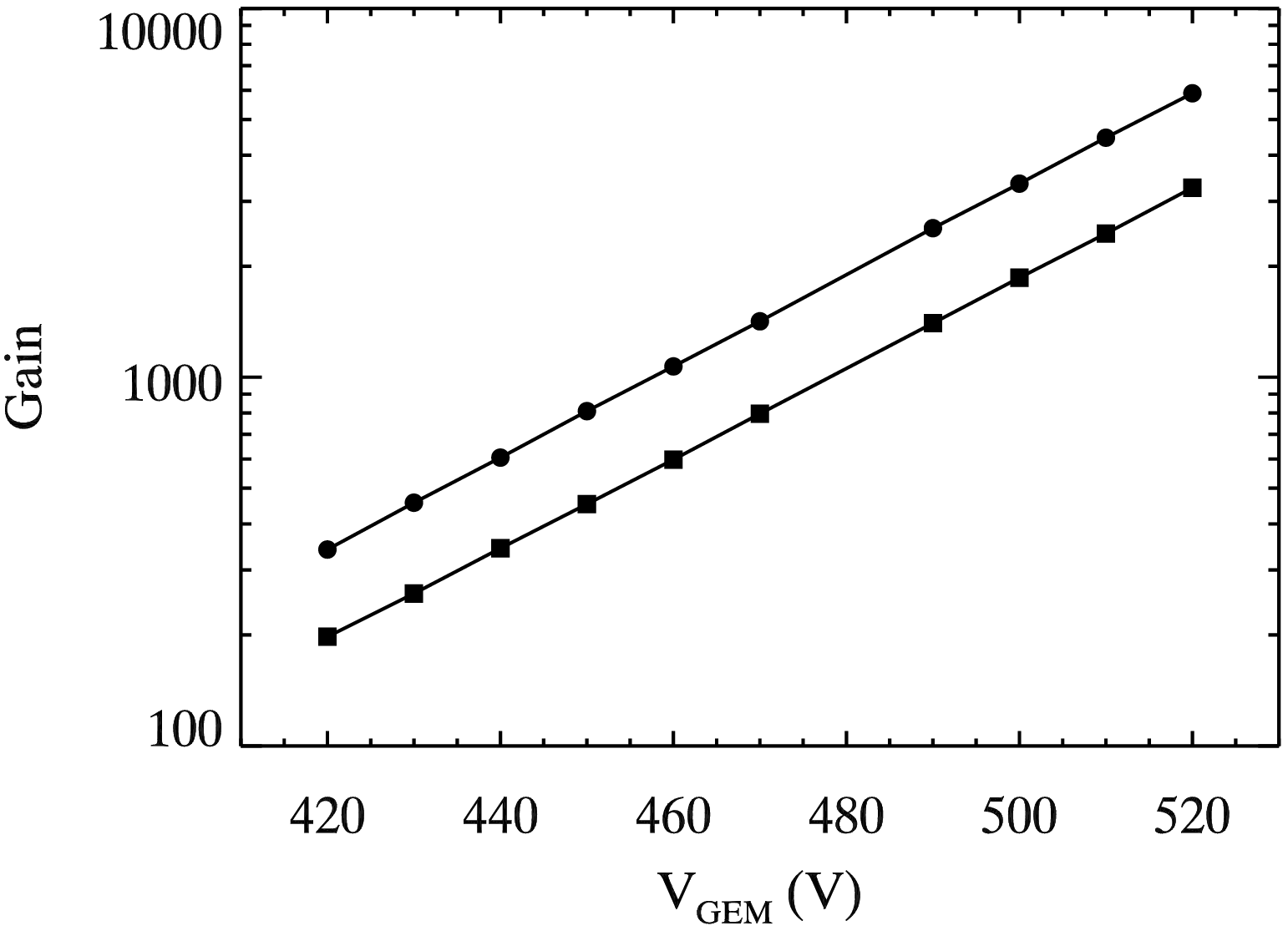}
\includegraphics[width=0.45\textwidth]{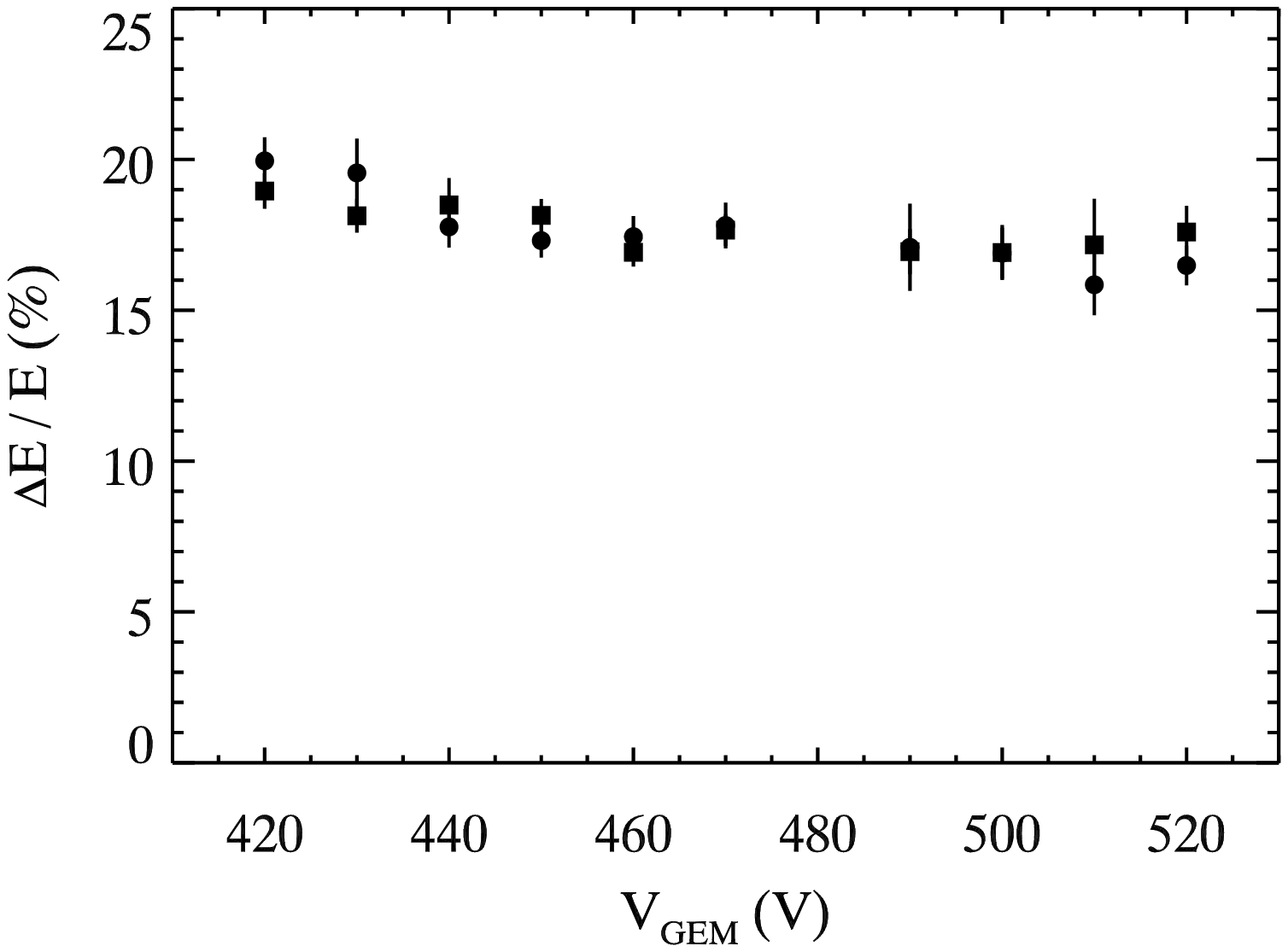}
\end{center}
\caption[gain]
{\label{fig:gain} 
The effective gain of the GEM and energy resolution versus GEM voltage. The circles indicate results measured from the GEM top layer while the squares are for measurements from the readout board. }
\end{figure}

The energy spectrum at 6.1 keV is shown in Figure~\ref{fig:spec}. We adopted a Gaussian profile to fit the spectrum. The fractional energy resolution ($\Delta E / E$) in full width at half maximum was found to be 17\%. Figure~\ref{fig:gain} shows the GEM gain and the energy resolution as a function of high voltage across the GEM. The drift field was 200 V/cm and the collecting field was 4 kV/cm.  To measure the electron track under the current electronics noise, an effective gain of 2000 is required and can be obtained by applying a high voltage of about 500 V across the GEM. The  energy resolution becomes constant when the GEM voltage is higher than 460 V. 

\section{Sensitivity}
\label{sec:sen}

The sensitivity of a polarimeter can be characterized by the minimum detectable polarization (MDP), which is defined as \cite{Weisskopf2010}
\begin{equation}
\label{modul}
    {\rm MDP} = \frac{4.29}{\mu \eta S A} \sqrt{\frac{\eta S A + B}{T}}
\end{equation}
at a confidence level of 99\%, where $S$ is the source rate in front of the detector (photons~cm$^{-2}$~s$^{-1}$), $B$ is the background rate in the detector (counts~s$^{-1}$), $T$ is the observing time in seconds, $A$ is the collecting area in cm$^2$, and $\eta$ is the detection efficiency.

\begin{figure}[!htbp]
\begin{center}
\includegraphics[width=0.6\textwidth]{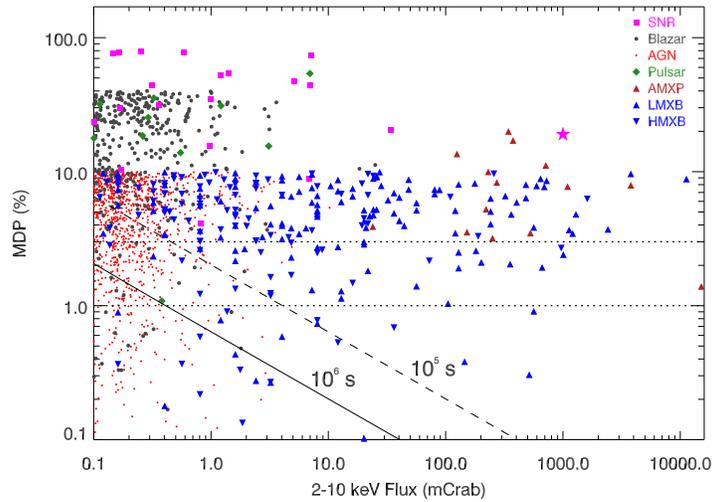}
\end{center}
\caption[mdp]
{\label{fig:mdp} 
MDP versus 2-10~keV X-ray flux. The dashed line and the solid line represent the MDPs with exposures of $10^5$~s and $10^6$~s, respectively. The dotted line indicates a systematic errors of 1\% or 3\%. Sources of different classes are assigned with a polarization degree randomly between 0 and the maximum possible value (see texts). Objects above the lines can be detected with XTP for poalrimetry.}
\end{figure}

One of the possible configurations for XTP is to mount 3 polarimeters behind 3 sets of high energy focusing telescopes, with a total effective area of 2400~cm$^2$. The internal background is estimated to be negligible, and the cosmic X-ray background (CXB) is considered to be the major component. We adopt the CXB spectrum measured with XMM-Newton \cite{Lumb2002}. Figure~\ref{fig:mdp} shows the MDP for XTP as a function of the 2--10~keV X-ray flux. The solid line indicates the MDP with an exposure of $10^6$~s and the dashed line is for an exposure of $10^5$~s. The dotted line corresponds to a limit of 1\% or 3\%, respectively,  depending on how good the systematics is controlled, suggesting that the sensitivity cannot be better than this level no matter how bright the source is. The symbols are X-ray objects of different classes, whose fluxes were estimated from the ROSAT All-Sky Survey Bright Source Catalog \cite{Voges1999}. The catalog provides the ROSAT PSPC count rates in the 0.1--2.4~keV band, which were translated into the 2--10 keV fluxes using the WebPIMMS tool assuming a power-law spectrum with a photon index of 2.0 subject to Galactic absorption with a column density of 10$^{21}$ cm$^{-2}$. For X-ray binaries, their 2--10~keV fluxes were adopted from the catalogs of Liu et al.\cite{Liu2000,Liu2001}. For each class of objects, we randomly assign a degree of polarization between 0 and the maximum possible value in theory: 80\% for supernova remnants \cite{Weisskopf1976,Weisskopf1978,Bykov2009}, 10\% for radio-quiet active galactic nuclei \cite{Goosmann2011}, 40\% for blazars \cite{Celotti1994}, 10\% for X-ray binaries in the hard state \cite{Schnittman2010}, 80\% for rotation-powered pulsars, and 20\% for accretion-powered millisecond X-ray pulsars \cite{Viironen2004}. This plot gives a sense about how many objects can be measured for polarimetry with XTP under current configuration.  We estimate that XTP is able to make a positive polarization measurement for about 1000 objects brighter than 0.1 mCrab with an exposure time of $10^5$~s and a systematics of 3\%.

\section{Conclusion}
\label{sec:con}

Here we report a new design for the TPC polarimeter, which has an extra window allowing high energy photons to leak through the detector and reach the focal plane, if the polarimeter is mounted above it.  This design can compensate the low efficiency of gaseous detectors and maximize the utilization of the focal plane of multilayer optics.  It is expected to have both high sensitivity in X-ray polarimetry and high transmission for hard X-rays. Such polarimeters are best suited for objects with high degrees of polarization, otherwise, imaging polarimeters which have an intrinsically low systematics are more suitable. For XTP, which is not a dedicated polarimetry mission, X-ray polarimetry provides new constraints to physical models in addition to timing and spectroscopy, and the scientific return of the mission is maximized only if the three means are all employed.  The prototype is tested to be functional. Calibrations for the modulation factors at different energies and the control of systematics can be done when a full scale readout electronics is available.


\end{document}